\documentclass[a4paper,12pt]{article}
\usepackage{amsmath, amssymb, latexsym, amscd, amsthm,amsfonts,amstext}
\usepackage[mathscr]{eucal}
\usepackage{graphicx}
\usepackage{subfig}

\numberwithin{equation}{section}
\input{tcilatex}
\begin{document}

\title{An inverse problem without the phase information}
\author{Michael V. Klibanov\thanks{
Department of Mathematics and Statistics, University of North Carolina at
Charlotte, Charlotte, NC 28223, USA; \texttt{mklibanv@uncc.edu}. } \thanks{%
This work was supported by US Army Research Laboratory and US Army Research
Office grant W911NF-15-1-0233 and by the Office of Naval Research grant
N00014-15-1-2330. }}
\date{}
\maketitle

\begin{abstract}
We prove a new uniqueness theorem for an inverse scattering problem without
the phase information for the 3-D Helmholtz equation. The spatially
distributed dielectric constant is the subject of the interest in this
problem. We consider the case when the modulus of the scattered wave field $%
\left\vert u_{sc}\right\vert $ is measured. The phase is not measured.
\end{abstract}

\graphicspath{
{FIGURES/}
 {pics/}}



\section{Introduction}

\label{sec:1}

Phaseless Inverse Scattering Problem (PISP) for a wave-like Partial
Differential Equation (PDE) with a complex valued solution is the problem of
the reconstruction of an unknown coefficient of this PDE from measurements
of the modulus of its solution on a certain set. The phase is not measured.
On the other hand, in conventional inverse scattering problems both the
modulus and the phase of the complex valued wave field are measured on
certain sets, see, e.g. \cite{Bao3,CK,Hu,Is,Li,Nov1,Nov2}.

Let $u=u_{0}+u_{sc}$ be the total wave field, where $u_{0}$ is the incident
wave field and $u_{sc}$ is the wave field scattered by a scatterer. The main
result of this paper is a uniqueness theorem for the PISP for the 3-D
Helmholtz equation in the case when the modulus $\left\vert
u_{sc}\right\vert $ of the scattered wave field is measured on a certain
surface. The closest previous publication is \cite{KlibHelm}. In \cite%
{KlibHelm} uniqueness was proven for the case when the modulus $\left\vert
u\right\vert $ of the total wave field is measured. Compared with \cite%
{KlibHelm}, the \emph{main difficulty} here is due to the interference of
two wave fields: $u$ and $u_{0}.$ To handle this difficulty, we develop here
some \emph{significantly new ideas}, which are presented in section 5 as
well as in (\ref{5.8})-(\ref{5.120}).

The first uniqueness result for a PISP was proven in \cite{KS} for the 1-D
case. In \cite{KSIAP,AML} uniqueness theorems were proven for the case when
the Schr\"{o}dinger equation 
\begin{equation}
\Delta v+k^{2}v-q\left( x\right) v=-\delta \left( x-y\right) ,x\in \mathbb{R}%
^{3}  \label{1.1}
\end{equation}%
was the underlying one. However, equation (\ref{1.1}) is easier to
investigate since the unknown coefficient $q\left( x\right) $ is not
multiplied by $k^{2}$ here, unlike the Helmholtz equation. Hence, unlike
both the current paper and \cite{KlibHelm}, it was not necessary to use in 
\cite{KSIAP,AML} the apparatus of the Riemannian geometry. Reconstruction
procedures for PISPs were developed for the Helmholtz equation and for the
Schr\"{o}dinger equation in \cite{KR1,KR2} and \cite{KR} respectively. In 
\cite{KNP} a modified reconstruction procedure of \cite{KR1} was numerically
implemented.

Our PISP is overdetermined, so as the one in \cite{KlibHelm} and some in 
\cite{KSIAP,AML}. Indeed, the data depend on six variables in our case,
whereas the unknown coefficient depends on three variables. On the other
hand, even if both the modulus and the phase of the wave field are measured,
uniqueness results in the 3-D case are unknown for the non-overdetermined
statements of inverse scattering problems. Still, the over-determination of
the data for some PISPs was not assumed in some uniqueness results. Those
are theorems 3 and 4 of \cite{KSIAP}, theorem 2 of \cite{AML} and theorems
of \cite{AA}. In particular, in \cite{AA} two PISPs for the Helmholtz
equation were considered. However, in each of the latter results the right
hand side of the corresponding PDE is a non-vanishing function $g\left(
x\right) $ rather than the $\delta -$function. This is because the last step
of the proof of each of those theorems is the application of the method of 
\cite{BukhKlib}, which is based on Carleman estimates, see, e.g. \cite%
{Ksurvey} for the recent survey of this method. However, the method of \cite%
{BukhKlib} does not work for the case when the $\delta -$function stands in
the right hand side of a PDE.

PISPs were also considered in \cite{Nov3,Nov4}. Statements of problems of
these works are different from ours, so as the results. These results
include both uniqueness theorems and reconstruction procedures.

Recall that a PISP is about the reconstruction of an unknown coefficient
from phaseless measurements. Along with PISPs, inverse problems of the
reconstruction of unknown surfaces of scatterers from phaseless data are
also of an obvious interest. In this regard, we refer to \cite%
{Am,Bao1,Bao2,Iv1,Iv2} for numerical solutions of some inverse scattering
problems without the phase information in the case when the surface of a
scatterer was reconstructed.

As to the applications of PISPs, they are in the lenseless imaging of
nanostructures. In the case when the size of a nanostructure is of the order
of 100 nanometers, which is 0.1 micron, the wavelength of the probing
radiation must be also about 0.1 micron. This corresponds to the frequency
of 2,997,924.58 Gigahertz, see, e.g. \cite{photonics}. It is well known that
it is impossible to measure the phase of an electromagnetic radiation at
such huge frequencies \cite{Dar,Pet,Ruhl}. Therefore, to image such a
nanostructure, one needs to compute its unknown spatially distributed
dielectric constant using measurements of only the intensity of the
scattered wave field (intensity is the square of the modulus). The second
application is in optical imaging of biological cells, since their sizes are
between 1 and 10 microns \cite{Phillips}.

In section 2 we formulate our PISP as well as the main result of our paper.
The rest of the paper is devoted to the proof of this result. In section 3
we explore a connection via the Fourier transform between our forward
problem for the Helmholtz equation and the Cauchy problem for a certain
hyperbolic equation. In section 4 we formulate three lemmata. In two lemmata
and one corollary of section 5 we present the above mentioned new ideas,
which handle the interference between the wave fields $u$\ and $u_{0}.$ In
section 6 we finalize the proof of the main result.

\section{Problem statement}

\label{sec:2}

Below $x=\left( x_{1},x_{2},x_{3}\right) \in \mathbb{R}^{3}.$ Consider a
non-magnetic and non-conductive medium, which occupies the whole space $%
\mathbb{R}^{3}.$ Let $c(x)$ be the spatially varying dielectric constant of
this medium. It was established in chapter 13 of the classical textbook of
Born and Wolf \cite{Born} that if the function $c(x)$ varies slowly enough
on the scales of the wavelength, then the scattering problem for Maxwell's
equations can be approximated by the scattering problem for the Helmholtz
equation for a certain component of the electric field. This justifies, from
the Physics standpoint, our work with the Helmholtz equation.

Let $\Omega ,\Psi ,G$ $\subset \mathbb{R}^{3}$ be three bounded domains and $%
\Omega \subset \Psi \subset G.$ Let $\partial \Psi =S\in C^{2}.$ Denote $%
2\rho =\min \left( dist\left( S,\partial \Omega \right) ,dist\left(
S,\partial G\right) \right) ,$ where \textquotedblleft $dist$" denotes the
Hausdorff distance. We assume that $\rho >0.$ For any number $\omega >0$ and
for every point $y\in \mathbb{R}^{3}$ denote $P_{\omega }\left( y\right)
=\left\{ x\in \mathbb{R}^{3}:\left\vert x-y\right\vert <\omega \right\} $.
We impose throughout the paper the following conditions on the function $%
c(x) $: 
\begin{equation}
c\in C^{15}(\mathbb{R}^{3}),  \label{2.101}
\end{equation}%
\begin{equation}
c(x)\geq 1\text{ in }\mathbb{R}^{3},  \label{2.102}
\end{equation}%
\begin{equation}
c(x)\geq 1+2\beta \text{ in }\Psi ,\beta =const.>0,  \label{2.103}
\end{equation}%
\begin{equation}
c\left( x\right) =1\quad \text{for }\>x\in \mathbb{R}^{3}\setminus G.
\label{2.104}
\end{equation}%
Inequality (\ref{2.102}) means that the dielectric constant of the medium is
not less than the dielectric constant of the vacuum, which is 1. So, (\ref%
{2.104}) means that the domain $G$ is embedded in the vacuum. In section 3
we use the fundamental solution of a hyperbolic equation with the
coefficient $c(x)$ in the principal part of its operator. The construction
of this solution works only if $c\in C^{15}(\mathbb{R}^{3})$ \cite{KR1,R3}.
In addition, the constructions of \cite{KR1,R3} require the regularity of
geodesic lines, see Condition below. We also note that the minimal
smoothness requirements for unknown coefficients are rarely a significant
concern in uniqueness theorems for multidimensional coefficient inverse
problems, see, e.g. \cite{Nov1,Nov2}, theorem 4.1 in Chapter 4 of \cite{R2}
and \cite{Ksurvey}.

The function $c(x)$ generates the conformal Riemannian metric, 
\begin{equation}
d\tau =\sqrt{c\left( x\right) }\left\vert dx\right\vert ,|dx|=\sqrt{%
(dx_{1})^{2}+(dx_{2})^{2}+(dx_{3})^{2}}.  \label{2.105}
\end{equation}%
\ We assume throughout the paper that the following condition holds:

\textbf{Condition}. \emph{Geodesic lines generated by the metric (\ref{2.105}%
) are regular. In other words, each pair of points }$x,y\in \mathbb{R}^{3}$%
\emph{\ can be connected by a single geodesic line }$\Gamma \left(
x,y\right) $\emph{.}

A sufficient condition for the regularity of geodesic lines was derived in 
\cite{R4}. For an arbitrary pair of points $x,y\in \mathbb{R}^{3}$ consider
the travel time $\tau (x,y)$ between them due to the Riemannian metric (\ref%
{2.105}). Then \cite{KR1,R2} 
\begin{equation}
|\nabla _{x}\tau (x,y)|^{2}=c(x),\quad  \label{2.106}
\end{equation}%
\begin{equation}
\tau (x,y)=O\left( \left\vert x-y\right\vert \right) \text{ }\>\mathrm{as}%
\text{ }\>x\rightarrow y.  \label{2.107}
\end{equation}%
The solution of the problem (\ref{2.106}), (\ref{2.107}) is \cite{KR1,R2} 
\begin{equation}
\tau \left( x,y\right) =\dint\limits_{\Gamma \left( x,y\right) }\sqrt{%
c\left( \xi \right) }d\sigma ,  \label{2.108}
\end{equation}%
where $d\sigma $ is the euclidean arc length. Using the above Condition, we
conclude that $\tau (x,y)$ is a single-valued function in $\mathbb{R}%
^{3}\times \mathbb{R}^{3}$.

Let $y\in \mathbb{R}^{3}$ be the position of the point source, $r=\left\vert
x-y\right\vert $ and $k>0$ be the wavenumber. We consider the Helmholtz
equation with the radiation condition at the infinity 
\begin{equation}
\Delta u+k^{2}c(x)u=-\delta (x-y),\quad x\in \mathbb{R}^{3},  \label{2.109}
\end{equation}%
\begin{equation}
\partial _{r}u-iku=o\left( 1/r\right) \>,r\rightarrow \infty .  \label{2.110}
\end{equation}%
Theorem 8.7 of \cite{CK} implies that for each $k>0$ the problem (\ref{2.109}%
), (\ref{2.110}) has unique solution $u\in C^{2}\left( \left\vert
x-y\right\vert \geq \varepsilon \right) ,\forall \varepsilon >0.$ Consider
the incident spherical wave $u_{0}$ and the scattered wave $u_{sc},$%
\begin{equation}
u_{0}\left( x,y,k\right) =\frac{\exp \left( ik\left\vert x-y\right\vert
\right) }{4\pi \left\vert x-y\right\vert },  \label{2.1100}
\end{equation}%
\begin{equation}
u_{sc}\left( x,y,k\right) =u\left( x,y,k\right) -u_{0}\left( x,y,k\right) .
\label{2.111}
\end{equation}

In this paper we consider the following PISP:

\textbf{Phaseless Inverse Scattering Problem (PISP)}. \emph{Assume that the
function }$c\left( x\right) $\emph{\ is given for }$x\in \mathbb{R}%
^{3}\diagdown \Omega $ \emph{and it is} \emph{unknown for }$x\in \Omega $%
\emph{. Suppose that the following function }$F\left( x,y,k\right) $\emph{\
is known}%
\begin{equation}
F\left( x,y,k\right) =\left\vert u_{sc}\left( x,y,k\right) \right\vert
,\forall y\in \partial S,\forall x\in P_{\rho }\left( y\right) ,x\neq
y,\forall k\in \left( a,b\right) ,  \label{2.112}
\end{equation}%
\emph{where }$\left( a,b\right) \subset \mathbb{R}_{+}=\left\{ k:k>0\right\} 
$\emph{\ is a certain interval. Determine the function }$c\left( x\right) $%
\emph{\ for }$x\in \Omega .$

The main result of this paper is Theorem 1:

\textbf{Theorem 1}. \emph{Assume that (\ref{2.101})- (\ref{2.104}), (\ref%
{2.111}), (\ref{2.112}) and Condition hold. Then the PISP has at most one
solution.}

The rest of this paper is devoted to the proof of this theorem. We assume
below that its conditions hold.

\section{Connection with a hyperbolic equation}

\label{sec:3}

As in \cite{KlibHelm,KR1}, consider the following Cauchy problem%
\begin{equation}
c\left( x\right) U_{tt}=\Delta U+\delta \left( x-y,t\right) ,x\in \mathbb{R}%
^{3},t>0,  \label{3.1}
\end{equation}%
\begin{equation}
U\mid _{t<0}\equiv 0.  \label{3.2}
\end{equation}%
For an arbitrary $T>0$ define the domains $K\left( y,T\right) $ and $K^{\ast
}\left( y,T\right) $ as 
\begin{equation*}
K(y,T)=\{(x,t):0<t\leq T-\tau (x,y)\},
\end{equation*}%
\begin{equation*}
K^{\ast }\left( y,T\right) =\{(x,t):\tau (x,y)\leq t\leq T-\tau (x,y)\}.
\end{equation*}%
Let $H\left( t\right) $ be the Heaviside function,%
\begin{equation*}
H\left( t\right) =\left\{ 
\begin{array}{c}
1,t>0, \\ 
0,t<0.%
\end{array}%
\right.
\end{equation*}

\textbf{Lemma 3.1} \cite{KR1}. \emph{For any fixed source position }$y\in 
\mathbb{R}^{3}$\emph{\ and for any }$T>0,$\emph{\ there exists unique
solution }$U(x,y,t)$\emph{\ of the problem (\ref{3.1}), (\ref{3.2}), which
can be represented in the domain }$K(y,T)$\emph{\ in the form}%
\begin{equation}
U(x,y,t)=A(x,y)\delta (t-\tau (x,y))+\widetilde{U}(x,y,t)H(t-\tau (x,y)),
\label{3.3}
\end{equation}%
\emph{where the function }$\widetilde{U}(x,y,t)\in C^{2}\left( K^{\ast
}\left( y,T\right) \right) $\emph{\ and }$A(x,y)$ \emph{is a certain
function such that }$A(x,y)>0,\forall x\in \mathbb{R}^{3},x\neq y$\emph{\
and }$A(x,y)$\emph{\ is continuous with respect to }$x,y$\emph{\ for }$x\neq
y.$\emph{\ Furthermore, for any bounded domain }$D\subset \mathbb{R}^{3}$%
\emph{\ the function }$U(x,y,t)$\emph{\ decays exponentially with respect to 
}$t$\emph{\ together with its }$x,t$ \emph{derivatives up to the second
order. In other words, there exist numbers }$\gamma =\gamma \left(
D,c,y\right) >0,Y=Y\left( D,c,y\right) >0$\emph{, }$t_{0}=t_{0}\left(
D,c,y\right) >0$\emph{\ depending only on listed parameters such that}%
\begin{equation}
\left\vert D_{x,t}^{\alpha }U\left( x,y,t\right) \right\vert \leq
Ye^{-\gamma \,t},\forall t\geq t_{0}\>,\forall \>x\in \overline{D}.
\label{3.4}
\end{equation}%
\emph{In (\ref{3.4}) }$\alpha =\left( \alpha _{1},\alpha _{2},\alpha
_{3},\alpha _{4}\right) $\emph{\ is the multi-index with non-negative
integer coordinates and }$\left\vert \alpha \right\vert =\alpha _{1}+\alpha
_{2}+\alpha _{3}+\alpha _{4}\leq 2.$\emph{\ Consider the Fourier transform }$%
F\left( U\right) $\emph{\ with respect to }$t$ \emph{of the function }$U$%
\emph{,} 
\begin{equation}
\mathcal{F}\left( U\right) \left( x,y,k\right) =\dint\limits_{0}^{\infty
}U\left( x,y,t\right) e^{ikt}dt:=V(x,y,k).  \label{3.5}
\end{equation}%
\emph{Then }%
\begin{equation}
V(x,y,k)=u(x,y,k),  \label{3.6}
\end{equation}%
\emph{\ where the function }$u(x,y,k)$\emph{\ is the solution of the problem
(\ref{2.109}), (\ref{2.110}).}

We note that the assertions of this lemma about both the exponential decay
and (\ref{3.6}) were derived in \cite{KR1} from the results of \cite{V1,V}.

For an arbitrary number $\theta >0$ denote 
\begin{equation*}
\mathbb{C}_{\theta }=\left\{ z\in \mathbb{C}:\func{Im}z>-\theta \right\} ,%
\mathbb{C}_{+}=\left\{ z\in \mathbb{C}:\func{Im}z>0\right\} .
\end{equation*}%
Lemma 3.2 follows immediately from (\ref{3.4})-(\ref{3.6}).

\textbf{Lemma 3.2 }\cite{KlibHelm}.\emph{\ For every }$x\in G$\emph{\ the
function }$u\left( x,y,k\right) $\emph{\ is analytic with respect to the
real variable }$k\in \mathbb{R}_{+}.$ \emph{Furthermore, the function }$%
u\left( x,y,k\right) $\emph{\ can be analytically continued with respect to }%
$k$\emph{\ from }$\mathbb{R}_{+}$\emph{\ in the half complex plane }$\mathbb{%
C}_{\gamma },$\emph{\ where }$\gamma =\gamma \left( G,c,y\right) >0$\emph{\
is the number of Lemma 3.1.}

Lemma 3.3 follows immediately from (\ref{3.3})-(\ref{3.6}).

\textbf{Lemma 3.3 }\cite{KlibHelm}. \emph{Let }$A(x,y)$\emph{\ be the
function in (\ref{3.3}). The asymptotic behavior of the function }$u(x,y,k)$%
\emph{\ is}%
\begin{equation}
u(x,y,k)=A(x,y)e^{ik\tau \left( x,y\right) }\left( 1+O\left( \frac{1}{k}%
\right) \right) ,\left\vert k\right\vert \rightarrow \infty ,k\in \mathbb{C}%
_{\gamma /2},\text{ }x\in \overline{G},  \label{3.7}
\end{equation}%
\emph{where}%
\begin{equation}
\left\vert O\left( \frac{1}{k}\right) \right\vert \leq \frac{M}{\left\vert
k\right\vert +1},k\in \mathbb{C}_{\gamma /2},\text{ }x\in \overline{G},
\label{3.8}
\end{equation}%
\emph{where the number }$M=M\left( G,c,y\right) >0$\emph{\ depends only on
listed parameters}.

\section{Three lemmata}

\label{sec:4}

\textbf{Lemma 4.1 }\cite{KlibHelm}. \emph{Let the function }$f\left(
k\right) $\emph{\ be analytic for all }$k\in \mathbb{R}.$ \emph{Then the
function }$\left\vert f\left( k\right) \right\vert $\emph{\ can be uniquely
determined for all }$k\in \mathbb{R}$ \emph{by the values of }$\left\vert
f\left( k\right) \right\vert $ \emph{\ for }$k\in \left( a,b\right) $\emph{.}

\textbf{Lemma 4.2 }\cite{KlibHelm}. \emph{Let }$\left\{ p_{j_{1}}\right\}
_{j_{1}=1}^{N_{1}}$\emph{\ and }$\left\{ q_{j_{2}}\right\}
_{j_{2}=1}^{N_{2}} $ \emph{be two sets of integers, all of which are
non-negative.\ In addition, consider two sets of complex numbers }$\left\{
\lambda _{j_{1}}\right\} _{j_{1}=1}^{N_{1}},\left\{ \theta _{j_{2}}\right\}
_{j_{2}=1}^{N_{2}}\subset C_{+}.$\emph{\ Assume that there exist two sets of
non-zero numbers }$\left\{ l_{j_{1}}\right\} _{j_{1}=1}^{N_{1}}$\emph{, }$%
\left\{ s_{j_{2}}\right\} _{j_{2}=1}^{N_{2}}\subset C$\emph{\ such that}%
\begin{equation}
\dsum\limits_{j_{1}=1}^{N_{1}}l_{j_{1}}t^{p_{j_{1}}}\exp \left( -i\overline{%
\lambda }_{j_{1}}t\right)
=\dsum\limits_{j_{1}=1}^{N_{1}}s_{j_{2}}t^{q_{j_{2}}}\exp \left( -i\overline{%
\theta }_{j_{2}}t\right) ,\forall t>0.  \label{4.1}
\end{equation}%
\emph{Then }$N_{1}=N_{2}=N$\emph{\ and numbers involved in (\ref{4.1}) can
be re-numbered in such a way that }%
\begin{equation*}
l_{j}=s_{j},p_{j}=q_{j},\lambda _{j}=\theta _{j},\forall j=1,...,N.
\end{equation*}

Lemma 4.3 follows immediately from Proposition 4.3 of \cite{Kl11}.

\textbf{Lemma 4.3.} \emph{Let }$f\left( k\right) $\emph{\ be an analytic
function in the half plane }$\mathbb{C}_{\gamma }.$\emph{\ Assume that the
function }$f\left( k\right) $\emph{\ has no zeros in }$\mathbb{C}_{+}\cup 
\mathbb{R}.$\emph{\ Also, let the asymptotic behavior of the function }$%
f\left( k\right) $ \emph{be:} \emph{\ } 
\begin{equation*}
f\left( k\right) =\frac{C}{k^{n}}\left[ 1+o\left( 1\right) +p_{1}\exp \left(
ikL_{1}\right) \right] \exp \left( ikL\right) ,\left\vert k\right\vert
\rightarrow \infty ,k\in \mathbb{C}_{+}\cup \mathbb{R},
\end{equation*}%
\emph{where }$C,p_{1}\in \mathbb{C}$\emph{, }$L\in \mathbb{R}$\emph{\ are
some numbers, }$L_{1}>0,$\emph{\ }$n\geq 0$\emph{\ and}%
\begin{equation}
\left\vert p_{1}\right\vert <1.  \label{4.2}
\end{equation}%
\emph{\ Then the values of }$\left\vert f\left( k\right) \right\vert $\emph{%
\ for }$k\in \mathbb{R}$\emph{\ uniquely determine the function }$f\left(
k\right) $ \emph{for} $k\in \mathbb{C}_{+}\cup \mathbb{R}.$

\section{Handling the interference of wave fields $u$ and $u_{0}$}

\label{sec:5}

The \emph{main difficulties} in the proofs of this section are caused by the
above mentioned interference of wave fields $u$ and $u_{0}.$

\textbf{Lemma 5.1}. \emph{Fix the point }$y\in S.$\emph{\ Then there exists
a sufficiently small number }$\omega _{0}=\omega _{0}\left( c,\beta \right)
\in \left( 0,\rho \right) $\emph{\ depending only on the function }$c$\emph{%
\ and the number }$\beta $\emph{, such that the asymptotic behavior of the
function }$u_{sc}(x,y,k)$\emph{\ for }$\left\vert k\right\vert \rightarrow
\infty ,k\in \mathbb{C}_{+}\cup \mathbb{R},x\in P_{\omega _{0}}\left(
y\right) $\emph{\ is }%
\begin{equation}
u_{sc}(x,y,k)=-\frac{\exp \left( ik\left\vert x-y\right\vert \right) }{4\pi
\left\vert x-y\right\vert }\left[ 1-B\left( x,y,k\right) +O\left( \frac{1}{k}%
\right) \right] ,  \label{3.9}
\end{equation}%
\emph{where the function }$B(x,y,k)$ \emph{is }%
\begin{equation}
B(x,y,k)=B_{1}(x,y)\exp \left[ ik\left( \tau \left( x,y\right) -\left\vert
x-y\right\vert \right) \right] ,  \label{3.90}
\end{equation}%
\begin{equation}
B_{1}(x,y)=4\pi \left\vert x-y\right\vert A(x,y).  \label{3.91}
\end{equation}%
\emph{The functions }$B_{1}(x,y)$ and $\left\vert 1-B(x,y,k)\right\vert $%
\emph{\ can be estimated as}%
\begin{equation}
B_{1}(x,y)\leq \frac{1+\beta /2}{1+\beta },\forall x\in P_{\omega
_{0}}\left( y\right) ,  \label{3.100}
\end{equation}%
\begin{equation}
\left\vert 1-B(x,y,k)\right\vert \geq \frac{\beta }{2\left( 1+\beta \right) }%
,\forall k\in \mathbb{C}_{+}\cup \mathbb{R},\forall x\in P_{\omega
_{0}}\left( y\right) .  \label{3.10}
\end{equation}%
\emph{\ Furthermore, there exists a sufficiently large number }$%
K_{0}=K_{0}\left( c,\beta \right) >0$ \emph{depending only on }$c$\emph{\
and }$\beta $ \emph{such that the following estimate holds}%
\begin{equation}
\left\vert u_{sc}(x,y,k)\right\vert \geq \frac{1}{4\pi \left\vert
x-y\right\vert }\cdot \frac{\beta }{4\left( 1+\beta \right) },\forall k\in
\left( \mathbb{C}_{+}\cup \mathbb{R}\right) \cap \left\{ \left\vert
k\right\vert \geq K_{0}\right\} ,\forall x\in P_{\omega _{0}}\left( y\right)
.  \label{3.11}
\end{equation}

\textbf{Proof}. Formulas (\ref{3.9})-(\ref{3.91}) follow immediately from
formulas (\ref{2.111}), (\ref{3.7}) and (\ref{3.8}). However, an \emph{%
essentially new} element here are estimates (\ref{3.100})-(\ref{3.11}).

By the formula (3.9) of \cite{KR1} the function $A(x,y)$ has the form%
\begin{equation}
A(x,y)=\frac{\sqrt{J\left( x,y\right) }}{4\pi \sqrt{c\left( x\right) }\tau
\left( x,y\right) },  \label{3.12}
\end{equation}%
where $J\left( x,y\right) >0$ is a certain function, which is continuous
with respect to $x,y\in \mathbb{R}^{3}.$ Also, formula (3.7) of \cite{KR1}
implies that $J\left( y,y\right) =1.$ Hence, there exists a sufficiently
small number $\omega _{1}\in \left( 0,\rho \right) $ such that 
\begin{equation}
\sqrt{J\left( x,y\right) }<1+\frac{\beta }{2},\forall x\in P_{\omega
_{1}}\left( y\right) .  \label{3.13}
\end{equation}%
Since $S=\partial \Psi ,$ then by (\ref{2.103}) there exists a sufficiently
small number $\omega _{2}\in \left( 0,\rho \right) $ such that%
\begin{equation}
c\left( x\right) \geq 1+\beta ,\forall x\in P_{\omega _{2}}\left( y\right) .
\label{3.14}
\end{equation}%
Let $\omega _{0}=\min \left( \omega _{1},\omega _{2}\right) .$ Then (\ref%
{2.108}) and (\ref{3.14}) imply that 
\begin{equation}
\tau \left( x,y\right) \geq \sqrt{1+\beta }\left\vert x-y\right\vert
,\forall x\in P_{\omega _{0}}\left( y\right) .  \label{3.15}
\end{equation}%
Hence, by (\ref{3.12})-(\ref{3.15})%
\begin{equation}
A(x,y)\leq \frac{1+\beta /2}{\left( 1+\beta \right) 4\pi \left\vert
x-y\right\vert },\forall x\in P_{\omega _{0}}\left( y\right) .  \label{3.16}
\end{equation}%
Hence, using (\ref{3.91}), we obtain (\ref{3.100}).

We now prove (\ref{3.10}). By (\ref{3.15})%
\begin{equation*}
\tau \left( x,y\right) -\left\vert x-y\right\vert \geq \left( \sqrt{1+\beta }%
-1\right) \left\vert x-y\right\vert \geq 0,\forall x\in P_{\omega
_{0}}\left( y\right) .
\end{equation*}%
Hence, 
\begin{equation*}
\left\vert \exp \left[ ik\left( \tau \left( x,y\right) -\left\vert
x-y\right\vert \right) \right] \right\vert \leq 1,\forall k\in \mathbb{C}%
_{+}\cup \mathbb{R},\forall x\in P_{\omega _{0}}\left( y\right) .
\end{equation*}%
Hence, using (\ref{3.90}) and (\ref{3.100}), we obtain 
\begin{equation*}
\left\vert B(x,y,k)\right\vert \leq \frac{1+\beta /2}{1+\beta },\forall k\in 
\mathbb{C}_{+}\cup \mathbb{R},\forall x\in P_{\omega _{0}}\left( y\right) .
\end{equation*}%
Hence,%
\begin{equation*}
\left\vert 1-B(x,y,k)\right\vert \geq 1-\left\vert B(x,y,k)\right\vert \geq 
\frac{\beta }{2\left( 1+\beta \right) },
\end{equation*}%
which proves (\ref{3.10}).

By (\ref{3.8}) one can choose a sufficiently large number $K_{0}=K_{0}\left(
c,\beta \right) >0$ such that in (\ref{3.9})%
\begin{equation*}
\left\vert O\left( \frac{1}{k}\right) \right\vert \leq \frac{\beta }{4\left(
1+\beta \right) },\forall k\in \left( \mathbb{C}_{+}\cup \mathbb{R}\right)
\cap \left\{ \left\vert k\right\vert \geq K_{0}\right\} ,\forall x\in
P_{\omega _{0}}\left( y\right) .
\end{equation*}%
This, (\ref{3.9}) and (\ref{3.10}) prove (\ref{3.11}). $\square $\newline

Everywhere below each complex/real valued zero of the function $%
u_{sc}(x,y,k):=\varphi _{x,y}\left( k\right) $ as the function of the
variable $k$ is counted as many times as its multiplicity is. For any number 
$z\in \mathbb{C}$ its complex conjugate is denoted as $\overline{z}.$

Lemmata 3.2 and 5.1 imply Corollary 5.1.

\textbf{Corollary 5.1}. \emph{Fix the point }$y\in S.$\emph{\ Then there
exists a sufficiently small number }$\omega _{0}=\omega _{0}\left( c,\beta
\right) \in \left( 0,\rho \right) $\emph{\ depending only on the function }$%
c $\emph{\ and the number }$\beta $\emph{\ such that for every fixed point }$%
x\in P_{\omega _{0}}\left( y\right) $ \emph{the function }$\varphi
_{x,y}\left( k\right) $ \emph{has at most a finite number of zeros in the
set }$\mathbb{C}_{+}\cup \mathbb{R}.$

\textbf{Lemma 5.2}.\emph{\ Fix a point }$y\in S$\emph{\ and a point }$x\in
P_{\omega _{0}}\left( y\right) ,$\emph{\ }$x\neq y,$ \emph{where }$\omega
_{0}$\emph{\ is the number of Lemma 5.1. Let }$\left\{ a_{j}\right\}
_{j=1}^{m_{1}}\subset \mathbb{R}$\emph{\ be the set of all real zeros of the
function }$\varphi _{x,y}\left( k\right) $ \emph{and }$\left\{ b_{j}\right\}
_{j=1}^{m_{1}}\subset \mathbb{C}_{+}$\emph{\ be the set of all those complex
zeros of }$\varphi _{x,y}\left( k\right) $, \emph{which are located in the
upper half complex plane }$\mathbb{C}_{+}$ \emph{(Corollary 5.1). Consider
the function }$\widetilde{\varphi }_{x,y}\left( k\right) $\emph{\ defined as}%
\begin{equation}
\widetilde{\varphi }_{x,y}\left( k\right) =\varphi _{x,y}\left( k\right)
\left( \prod\limits_{j=1}^{m_{1}}\frac{1}{k-a_{j}}\right) \left(
\prod\limits_{j=1}^{m_{2}}\frac{k-\overline{b}_{j}}{k-b_{j}}\right) .
\label{4.3}
\end{equation}%
\emph{Then the values of the modulus }$\left\vert \varphi _{x,y}\left(
k\right) \right\vert $\emph{\ for }$k\in \mathbb{R}$\emph{\ together with
the values of all real zeros} \emph{uniquely determine the function }$%
\widetilde{\varphi }_{x,y}\left( k\right) $\emph{\ for }$k\in \mathbb{R}.$

\textbf{Proof}. Note that 
\begin{equation*}
\left\vert \frac{k-\overline{z}}{k-z}\right\vert =1,\forall k\in \mathbb{R}%
,\forall z\in \mathbb{C}.
\end{equation*}%
Hence, 
\begin{equation}
\left\vert \widetilde{\varphi }_{x,y}\left( k\right) \right\vert =\left\vert
\varphi _{x,y}\left( k\right) \right\vert \prod\limits_{j=1}^{m_{1}}\frac{1}{%
\left\vert k-a_{j}\right\vert },\forall k\in \mathbb{R}.  \label{4.4}
\end{equation}

We now need to apply Lemma 4.3. To do this, we use Lemma 5.1. In notations
of Lemma 4.3%
\begin{equation*}
C=-\frac{1}{4\pi \left\vert x-y\right\vert },L=\left\vert x-y\right\vert ,
\end{equation*}%
\begin{equation*}
p_{1}=B_{1}(x,y)>0,L_{1}=\tau \left( x,y\right) -\left\vert x-y\right\vert
>0.
\end{equation*}%
It follows from (\ref{3.100}) that $p_{1}<1,$ which means that condition (%
\ref{4.2}) holds. Next, (\ref{4.3}) implies that the function $\widetilde{%
\varphi }_{x,y}\left( k\right) $ does not have zeros in $\mathbb{C}_{+}\cup 
\mathbb{R}.$ To finish this proof, we refer to Lemma 4.3 and (\ref{4.4}). $%
\square $

\section{Proof of Theorem 1}

\label{sec:6}

In this proof we use results of sections 3-5. The part of the proof which
handles the interference of wave fields $u$ and $u_{0}$ is from (\ref{5.8})
to (\ref{5.120}).

Assume that there exist two functions $c_{1},c_{2}$ which correspond to the
same function $F\left( x,y,k\right) $ in (\ref{2.112}). Then%
\begin{equation}
c_{1}\left( x\right) =c_{2}\left( x\right) ,x\in \mathbb{R}^{3}\diagdown
\Omega .  \label{6.1}
\end{equation}%
Our goal is to prove that 
\begin{equation}
c_{1}\left( x\right) =c_{2}\left( x\right) ,x\in \Omega .  \label{6.2}
\end{equation}

Let $u_{1}\left( x,y,k\right) $ and $u_{2}\left( x,y,k\right) $ be two
functions $u\left( x,y,k\right) $ which correspond to coefficients $c_{1}$
and $c_{2}$ respectively and let $u_{1,sc}\left( x,y,k\right) $ and $%
u_{2,sc}\left( x,y,k\right) $ be two corresponding functions $u_{sc}\left(
x,y,k\right) $. Fix a point $y\in S.$ Let $\omega ^{\ast }=\min \left(
\omega _{0}\left( c_{1},\beta \right) ,\omega _{0}\left( c_{2},\beta \right)
\right) ,$ where the number $\omega _{0}=\omega _{0}\left( c,\beta \right)
\in \left( 0,\rho \right) $ was defined in Lemma 5.1. Fix a point $x\in
P_{\omega ^{\ast }}\left( y\right) ,x\neq y.$ Denote%
\begin{equation}
\varphi _{1}\left( k\right) =u_{1,sc}\left( x,y,k\right) ,\varphi _{2}\left(
k\right) =u_{2,sc}\left( x,y,k\right) .  \label{5.1}
\end{equation}%
It follows from (\ref{2.112}) and Lemma 4.1 that 
\begin{equation}
\left\vert \varphi _{1}\left( k\right) \right\vert =\left\vert \varphi
_{2}\left( k\right) \right\vert ,\forall k\in \mathbb{R}.  \label{5.2}
\end{equation}%
Using (\ref{5.2}), we obtain, similarly with \cite{KlibHelm}, that real
zeros of functions $\varphi _{1}\left( k\right) $ and $\varphi _{2}\left(
k\right) $ coincide.

Consider now zeros of functions $\varphi _{1}\left( k\right) $ and $\varphi
_{2}\left( k\right) $ in the upper half complex plane $\mathbb{C}_{+}.$ By
Corollary 5.1 each of these functions has at most a finite number of such
zeros. Let $\left\{ d_{j}\right\} _{j=1}^{n_{1}}\subset \mathbb{C}_{+}$ and $%
\left\{ e_{j}\right\} _{j=1}^{n_{2}}\subset \mathbb{C}_{+}$ be zeros of
functions $\varphi _{1}\left( k\right) $ and $\varphi _{2}\left( k\right) $
respectively. Then (\ref{5.2}) and Lemma 5.2 imply that 
\begin{equation}
\varphi _{1}\left( k\right) \dprod\limits_{j=1}^{n_{1}}\frac{k-\overline{d}%
_{j}}{k-d_{j}}=\varphi _{2}\left( k\right) \dprod\limits_{j=1}^{n_{2}}\frac{%
k-\overline{e}_{j}}{k-e_{j}},\forall k\in \mathbb{R}.  \label{5.3}
\end{equation}%
Hence,%
\begin{equation*}
\varphi _{1}\left( k\right) \dprod\limits_{j=1}^{n_{2}}\frac{k-e_{j}}{k-%
\overline{e}_{j}}=\varphi _{2}\left( k\right) \dprod\limits_{j=1}^{n_{1}}%
\frac{k-d_{j}}{k-\overline{d}_{j}},\forall k\in \mathbb{R}.
\end{equation*}%
Hence, 
\begin{equation}
\varphi _{1}\left( k\right) +\varphi _{1}\left( k\right) \left(
\dprod\limits_{j=1}^{n_{2}}\frac{k-e_{j}}{k-\overline{e}_{j}}-1\right)
=\varphi _{2}\left( k\right) +\varphi _{2}\left( k\right) \left(
\dprod\limits_{j=1}^{n_{1}}\frac{k-d_{j}}{k-\overline{d}_{j}}-1\right)
,\forall k\in \mathbb{R}.  \label{5.4}
\end{equation}%
Denote 
\begin{equation*}
g_{1}\left( k\right) =\dprod\limits_{j=1}^{n_{1}}\frac{k-d_{j}}{k-\overline{d%
}_{j}}-1,\text{ }g_{2}\left( k\right) =\dprod\limits_{j=1}^{n_{2}}\frac{%
k-e_{j}}{k-\overline{e}_{j}}-1.
\end{equation*}%
Let the multiplicity of the zero $e_{j}$ be $p_{j}$ and the multiplicity of
the zero $d_{j}$ be $q_{j}.$ Then the partial fraction expansion implies
that there exist numbers $X_{j}$ and $Y_{j}$ such that 
\begin{equation}
g_{1}\left( k\right) =\dsum\limits_{j_{1}=1}^{n_{1}^{\prime }}\frac{X_{j_{1}}%
}{\left( k-\overline{d}_{j_{1}}\right) ^{p_{j_{1}}}},\text{ }g_{2}\left(
k\right) =\dsum\limits_{j_{2}=1}^{n_{2}^{\prime }}\frac{Y_{j_{2}}}{\left( k-%
\overline{e}_{j_{2}}\right) ^{q_{j_{2}}}},  \label{5.5}
\end{equation}%
In (\ref{5.5}) $n_{1}^{\prime },n_{2}^{\prime }$ are some positive integers
such that $n_{1}^{\prime }\leq n_{1},n_{2}^{\prime }\leq n_{2}.$ Applying
the inverse Fourier transform $\mathcal{F}^{-1}$ to functions $g_{1}\left(
k\right) $ and $g_{2}\left( k\right) $ in (\ref{5.4}), we obtain \cite%
{KlibHelm}%
\begin{equation}
\mathcal{F}^{-1}\left( g_{1}\right) =Q_{1}\left( t\right) =H\left( t\right)
\dsum\limits_{j_{1}=1}^{n_{1}^{\prime }}X_{j_{1}}\frac{\left( -1\right)
^{p_{j_{1}}-1}i^{p_{j_{1}}}}{\left( p_{j_{1}}-1\right) !}\cdot
t^{p_{j_{1}}-1}\exp \left( -i\overline{d}_{j_{1}}t\right) .  \label{5.6}
\end{equation}%
\begin{equation}
\mathcal{F}^{-1}\left( g_{2}\right) =Q_{2}\left( t\right) =H\left( t\right)
\dsum\limits_{j_{2}=1}^{n_{2}^{\prime }}Y_{j_{2}}\frac{\left( -1\right)
^{q_{j_{2}}-1}i^{q_{j_{2}}}}{\left( q_{j_{2}}-1\right) !}\cdot
t^{q_{j_{2}}-1}\exp \left( -i\overline{e}_{j_{2}}t\right) ,  \label{5.7}
\end{equation}%
By Lemma 3.1 and (\ref{5.1}) the inverse Fourier transform of each of
functions $\varphi _{1}\left( k\right) $ and $\varphi _{2}\left( k\right) $
exists and 
\begin{equation}
\mathcal{F}^{-1}\left( \varphi _{1}\right) =U_{1}\left( x,y,t\right) -\frac{%
\delta \left( t-\left\vert x-y\right\vert \right) }{4\pi \left\vert
x-y\right\vert },  \label{5.8}
\end{equation}%
\begin{equation}
\mathcal{F}^{-1}\left( \varphi _{2}\right) =U_{2}\left( x,y,t\right) -\frac{%
\delta \left( t-\left\vert x-y\right\vert \right) }{4\pi \left\vert
x-y\right\vert },  \label{5.9}
\end{equation}%
where functions $U_{1}\left( x,y,t\right) $ and $U_{2}\left( x,y,t\right) $
are solutions of the Cauchy problem (\ref{3.1}), (\ref{3.2}) with $c\left(
x\right) =c_{1}\left( x\right) $ and $c\left( x\right) =c_{2}\left( x\right) 
$ respectively. In (\ref{5.8}) and (\ref{5.9}) we have used the fact that 
\begin{equation*}
\mathcal{F}^{-1}\left( u_{0}\right) =\frac{\delta \left( t-\left\vert
x-y\right\vert \right) }{4\pi \left\vert x-y\right\vert }.
\end{equation*}

Therefore, using the convolution theorem and (\ref{5.4})- (\ref{5.9}), we
obtain%
\begin{equation}
U_{1}\left( x,y,t\right) +\dint\limits_{0}^{t}\left( U_{1}\left( x,y,t-\xi
\right) -\frac{\delta \left( t-\xi -\left\vert x-y\right\vert \right) }{4\pi
\left\vert x-y\right\vert }\right) Q_{2}\left( \xi \right) d\xi  \label{5.10}
\end{equation}%
\begin{equation*}
=U_{2}\left( x,y,t\right) +\dint\limits_{0}^{t}\left( U_{2}\left( x,y,t-\xi
\right) -\frac{\delta \left( t-\xi -\left\vert x-y\right\vert \right) }{4\pi
\left\vert x-y\right\vert }\right) Q_{1}\left( \xi \right) d\xi .
\end{equation*}%
Since $x\in P_{\omega ^{\ast }}\left( y\right) ,x\neq y$ and by (\ref{3.15}) 
$\tau \left( x,y\right) \geq \sqrt{1+\beta }\left\vert x-y\right\vert ,$
then (\ref{3.3}) implies that 
\begin{equation}
U_{1}\left( x,y,t\right) =U_{2}\left( x,y,t\right) =0\text{, }\forall t\in
\left( \left\vert x-y\right\vert ,\sqrt{1+\beta /2}\left\vert x-y\right\vert
\right) .  \label{5.11}
\end{equation}%
Hence, using (\ref{5.10}) and (\ref{5.11}), we obtain 
\begin{equation}
Q_{1}\left( t-\left\vert x-y\right\vert \right) =Q_{2}\left( t-\left\vert
x-y\right\vert \right) ,\forall t\in \left( \left\vert x-y\right\vert ,\sqrt{%
1+\beta /2}\left\vert x-y\right\vert \right) .  \label{5.12}
\end{equation}%
Since by (\ref{5.6}) and (\ref{5.7}) each of functions $Q_{1}\left( t\right) 
$ and $Q_{2}\left( t\right) $ is analytic as the function of the real
variable $t>0,$ then (\ref{5.12}) implies that 
\begin{equation}
Q_{1}\left( t\right) =Q_{2}\left( t\right) ,\forall t>0.  \label{5.120}
\end{equation}
Thus, using Lemma 4.2, (\ref{5.6}), (\ref{5.7}) and (\ref{5.120}), we obtain
that zeros of functions $\varphi _{1}\left( k\right) $ and $\varphi
_{2}\left( k\right) $ in $\mathbb{C}_{+}\cup \mathbb{R}$ coincide.

Hence, by (\ref{5.3}) $\varphi _{1}\left( k\right) =\varphi _{2}\left(
k\right) ,\forall k\in \mathbb{R}.$ Hence, (\ref{2.1100}), (\ref{2.111}) and
(\ref{5.1}) imply that 
\begin{equation}
u_{1}\left( x,y,k\right) =u_{2}\left( x,y,k\right) ,\forall k\in \mathbb{R}.
\label{5.13}
\end{equation}%
Since $y$ is an arbitrary point of the surface $S\subset \mathbb{R}%
^{3}\diagdown \Omega $, $x\neq y$ is an arbitrary point of the ball $%
P_{\omega ^{\ast }}\left( y\right) $ and $P_{\omega ^{\ast }}\left( y\right)
\cap \overline{\Omega }=\varnothing ,$ then, using (\ref{6.1}), (\ref{5.13})
and the well known theorem about the uniqueness of the continuation of the
solution of the elliptic equation of the second order (see, e.g. \cite{LRS}%
), we obtain 
\begin{equation*}
u_{1}\left( x,y,k\right) =u_{2}\left( x,y,k\right) ,\forall k\in \mathbb{R}%
,\forall y\in S,\forall x\in \mathbb{R}^{3}\diagdown \Psi .
\end{equation*}%
Hence, using Lemma 3.1 and the fact that the Fourier transform is
one-to-one, we obtain

\begin{equation}
U_{1}\left( x,y,t\right) =U_{2}\left( x,y,t\right) ,\forall t>0,\forall y\in
S,\forall x\in \mathbb{R}^{3}\diagdown \Psi .  \label{5.14}
\end{equation}

Thus, (\ref{3.3}) and (\ref{5.14}) imply that 
\begin{equation}
\tau _{1}\left( x,y\right) =\tau _{2}\left( x,y\right) ,\forall x,y\in S,
\label{5.15}
\end{equation}%
where functions $\tau _{1}\left( x,y\right) $ and $\tau _{2}\left(
x,y\right) $ correspond to the function $\tau \left( x,y\right) $ for $%
c=c_{1}$ and $c=c_{2}$ respectively.

As the last step of the proof, we now apply to (\ref{5.15}) theorem 3.4 of
Chapter 3 of the book \cite{R2}. We follow notations of that theorem. Let 
\begin{equation}
n_{1}\left( x\right) =\sqrt{c_{1}\left( x\right) }\text{ and }n_{2}\left(
x\right) =\sqrt{c_{2}\left( x\right) }.  \label{5.16}
\end{equation}%
By (\ref{2.102}) 
\begin{equation}
n_{1}\left( x\right) ,n_{2}\left( x\right) \geq 1.  \label{5.17}
\end{equation}%
Also, using (\ref{2.101}) and (\ref{2.104}) we obtain that there exists a
number $n_{00}>1$ such that 
\begin{equation}
\left\Vert n_{1}\left( x\right) \right\Vert _{C^{2}\left( \overline{\Psi }%
\right) },\left\Vert n_{2}\left( x\right) \right\Vert _{C^{2}\left( 
\overline{\Psi }\right) }\leq n_{00}.  \label{5.18}
\end{equation}%
Denote $\Lambda \left( 1,n_{00}\right) $ the class of functions $n\left(
x\right) $ such that the following two conditions hold for every function $%
n\left( x\right) \in \Lambda \left( 1,n_{00}\right) :$

\begin{enumerate}
\item  The function $c\left( x\right) =n^{2}\left( x\right) $ satisfies
conditions (\ref{2.101})-(\ref{2.104}) as well as Condition of section 2.

\item $\left\Vert n\left( x\right) \right\Vert _{C^{2}\left( \overline{\Psi }%
\right) }\leq n_{00}.$
\end{enumerate}

By (\ref{5.16})-(\ref{5.18}) both functions $n_{1}\left( x\right)
,n_{2}\left( x\right) \in \Lambda \left( 1,n_{00}\right) .$ Therefore, (\ref%
{6.1}), (\ref{5.15}) and the estimate (3.66) of theorem 3.4 of Chapter 3 of
the book \cite{R2} imply (\ref{6.2}). \ $\square $

\end{document}